# Resistance and lifetime measurements of polymer solar cells using glycerol doped poly[3,4-ethylenedioxythiophene]: poly[styrenesulfonate] hole injection layers

Emma Lewis[1], Bhaskar Mantha[2] and Richard P. Barber, Jr.[1]

[1]Department of Physics and Center for Nanostructures, Santa Clara University

[2]Department of Electrical Engineering, Santa Clara University

## Abstract

We have performed resistivity measurements of poly[3,4-ethylenedioxythiophene]: poly[styrenesulfonate] (PEDOT:PSS) films with varying concentrations of glycerol. Resistivity is seen to decrease exponentially from roughly 3 $\Omega$-cm for pure PEDOT:PSS to $3 \times 10^{-2}$ $\Omega$-cm for 35 mg/cm$^3$ glycerol in PEDOT:PSS. Beyond this concentration adding glycerol does not significantly change resistivity. Bulk heterojunction polymer solar cells using these variously doped PEDOT:PSS layers as electrodes were studied to characterize the effects on efficiency and lifetime. Although our data display significant scatter, lowering the resistance of the PEDOT:PSS layers results in lower device resistance and higher efficiency as expected. We also note that the lifetime of the devices tends to be reduced as the glycerol content of PEDOT:PSS is increased. Many devices show an initial increase in efficiency followed by a roughly exponential decay. This effect is explained based on concomitant changes in the zero bias conductance of the samples under dark conditions.

**Introduction**

Organic photovoltaics (OPVs) continue to attract interest as a low-cost and mechanically robust alternative to Si-based technologies. The primary challenges facing OPVs have been lower efficiency and lifetime, however significant progress has occurred over the last decade [1]. Furthermore, the primary advantage of OPVs, manufacturability, has been clearly demonstrated [2], [3]. A common approach to producing theses solar cells relies on the transparent conductive layer indium-tin-oxide (ITO). However the use of ITO imposes limits on the flexibility of the cells and hence the manufacturability[4], [5]. A potential replacement for the ITO layer is poly[3,4-ethylenedioxythiophene]: poly[styrenesulfonate] (PEDOT:PSS). PEDOT:PSS is already commonly used as a hole conducting planarizing layer over the ITO in many solar cell architectures [6]. To overcome the relatively low conductivity of PEDOT:PSS various studies have utilized additives [4], [5], [7]–[13] including glycerol[8], [10], [11] and carbon nanotubes[5]. The aging of the PEDOT:PSS electrodes has been investigated [4]. However, we are unaware of stability studies of actual solar cell devices based on these modified electrodes.

Given the importance of cell stability [1], [14], [15], the central motivation for this work is to find any effects that enhancing the PEDOT:PSS conductivity might have on device lifetime. Specifically, we report measurements of series resistance, efficiency and lifetime of solar cells using various glycerol/PEDOT:PSS (G-PEDOT:PSS) films as the transparent contact. These measurements are correlated with the resistivity of films cast from the same glycerol/PEDOT:PSS blends. In addition, previous work showed a dramatic decrease in the resistivity (2 orders of magnitude) of PEDOT:PSS when *30 mg* per *cm$^3$* was added [11]. We were also interested in whether this onset could be more finely tuned.

**Experimental**

A series of G-PEDOT:PSS (Aldrich 483095) solutions were prepared with glycerol concentrations of *0, 9.2, 22.3, 36.1, 70.7* and *95.0 mg* per *cm³* of PEDOT:PSS. Films were spin cast from these solutions and annealed at *200 °C* for two hours to produce electrical transport samples (Fig. 1a) or transparent contacts for solar cell measurements (Fig. 1b). All "wet" preparation, processing and annealing steps were conducted in an inert atmosphere glove box. Film resistance was measured using a variable current source with separate voltage leads to derive current voltage (*IV*) curves (contacts schematically shown in Fig. 1a). Film resistance (*R*) was derived from $R = \lim_{I \to 0} \frac{\partial V}{\partial I}$, however no significant nonlinearity was observed in any of these samples. Given that resistance is given by $R = \frac{\rho L}{Wd}$ where *L* is the length of the film, *W* is the width, *d* is the thickness and $\rho$ is the material's resistivity; we can also define the sheet resistance or resistance per square $R_{sq} = \frac{\rho}{d}$ since *L/W* is the number of squares. Finally, film thickness measurements obtained from a Gaertner L116B ellipsometer allow us to calculate the resistivity of our samples ($\rho = R_{sq} d$).

Solar cells were fabricated using a *0.16* mole fraction (equal weight) blend of [6,6]-Phenyl C61 butyric acid methyl ester (PCBM) in Poly(3-hexylthiophene-2,5-diyl) (P3HT) spin cast on to our G-PEDOT:PSS electrodes from a *1.5* weight percent solution in chlorobenzene. These samples were annealed for one hour at *190 °C* and then transferred to a bell-jar evaporator system equipped with a quartz crystal thickness monitor where they were finished by evaporating ~*1 nm* of LiF followed by *100 nm* of Al to form the top electrodes. The evaporator is not integrated into the glove box, so samples were transferred between the two using a vacuum tight vessel carrying dry nitrogen atmosphere. The elapsed time that samples were exposed to ambient air was typically under 5 minutes. Current-voltage (*IV*) characteristics of the devices were measured in ambient conditions alternately in darkness and illuminated by a PV Measurements, Inc. Small-Area Class-B Solar Simulator. Automated transport data collection utilized a MATLAB controlled routine via an IEEE 488 Bus interfaced Keithley 2400 SourceMeter. *IV* curves were typically acquired at 15 minute intervals. However as samples aged, this time was sometimes increased to 30 minutes, as changes became smaller.

**Results and Discussion**

We have plotted the sheet resistance (left axis) for samples of G-PEDOT:PSS as glycerol concentration is varied in Fig. 2; these values are in reasonable agreement with previous work[11]. We observe that for glycerol doping below about *30 mg/cm$^3$* the sheet resistance drops roughly exponentially with concentration, a more finely tuned result that was previously reported[11]. Beyond this value, there is little change. Overlaid on Fig. 2 are our resistivity values from the same data (right axis). The nearly point-to-point correspondence between the sheet resistance and resistivity indicates consistent film thicknesses for all of the samples. The averaged thickness for the sixteen samples was *340±39 nm*, with only three outlier samples (outside of the error bars).

Fig. 3 displays the results for three parameters of merit for the solar cell devices: power conversion efficiency $\eta$, series resistance $R_S$ and characteristic time $\tau$ (lifetime) as plotted against the sheet resistance $R_{sq}$ of the various G-PEDOT:PSS blends. The efficiencies are roughly two orders of magnitude below optimized PCBM/P3HT devices [16], however our focus is on the lifetime behavior and not the efficiency while using this ITO-free sample structure. It is important to note that we saw consistent device performance behavior throughout these measurements, and we believe that this reproducibility supports the validity of the results. We have so far been unable to find the cause of the poor device performance. We should note that in measurements of devices on the same substrate both with and without the ITO layer (but having the same PEDOT:PSS and active layer), we found that ITO-free samples performed as well or better than those with ITO. Series resistance is calculated by measuring the asymptotic behavior of the *IV* curve at the highest applied voltage. In order to measure $\tau$, *IV* curves were taken in ambient atmospheric conditions at *15 minute* intervals for up to *24 hours*. From these data we derived the power conversion efficiency as a function of time *η(t)* and fit those results to an exponential decay model to extract $\tau$ [17]. There is significant scatter in the results, however the trends are apparent (regression lines shown) and probably meaningful. As the transparent G-PEDOT:PSS resistance is lowered we see both an increase in efficiency and a decrease in $R_S$. These results are consistent given that $R_S$ directly affects the output current and hence the efficiency. Furthermore it is reasonable to expect that $R_S$ will depend in part on the resistivity of the PEDOT:PSS contact since that should contribute to the device resistance. In Fig. 3c we observe that decreasing the G-PEDOT:PSS resistance *reduces* the lifetime of the devices albeit only about a factor of two over two decades of resistance change. We also point out that most of the *IV* curves recorded are nearly linear (for example see Fig. 4), yielding fill factors which are very close to *0.25* as would be expected.

Fig. 4 shows a typical set of *IV* curves for an illuminated device as it ages (some traces have been removed for clarity). The arrow indicates the direction of increasing time. Besides the poor fill factors (~*0.25*) we note that both the initial short circuit current (current axis intercept) and the open circuit voltage (zero-current crossing) *increase* from the initial values. In other words, the device efficiency is initially increasing before a decay process begins. This kind of behavior has been observed in previous samples from our laboratory [18], albeit as a much smaller effect. A clue for understanding this behavior is exhibited in the un-illuminated *IV* data for the same sample as displayed in the inset of Fig. 4. These curves suggest a characteristic sample resistance which increases with time (again indicated by the arrow). We postulate that this initially "low" resistance represents a "leakage" or parallel conduction path that effectively shunts current directly through the device. This mechanism could potentially reduce both the current and voltage as long as the resistance is significantly low. To test this idea, we plot in Fig. 5 $\eta(t)$ for a sample which shows an initial increase in $\eta$ nearly two orders of magnitude before a decay mechanism becomes dominant. On the same figure (right axis) is the zero bias conductance for the same film taken in the dark. We note that the initial dark conductance drops precipitously at about *100* minutes, the same time that the maximum $\eta$ is observed. Beyond that time we again see the typical exponential decay of the device efficiency [17]. Although we cannot speculate on the exact mechanism of this effect, it is certainly enhanced for samples that utilize G-PEDOT:PSS contacts.

**Conclusions**

In summary, we have characterized the resistivity PEDOT:PSS films as a function of glycerol doping. We have also measured the efficiency, series resistance and lifetime of PCBM/P3HT solar cells utilizing these G-PEDOT:PSS films as the transparent contact. These parameters of merit were found to vary with the glycerol doping, with the lifetime of cells negatively impacted by the addition of glycerol. We have also observed that these solar cells often exhibit efficiencies which initially increase before an exponential decay begins. We have proposed a simple leakage mechanism for this behavior based on the zero-bias resistance of the devices.

**Acknowledgements**

We acknowledge both G. Laskowski and G. Sloan for invaluable technical assistance. Funding was provided in part by a grant from IntelliVision Technologies and a Santa Clara University IBM Faculty Research Grant.

**Figure Captions**

Fig. 1. a) Schematic sample layout for resistivity measurements and b) solar cell architecture.

Fig. 2. Sheet resistance (left axis) and resistivity (right axis) for G-PEDOT:PSS films as a function of glycerol concentration.

Fig. 3. Solar cell parameters of merit for PCBM/P3HT devices as dependent on the sheet resistance of the G-PEDOT:PSS films used as the transparent electrodes: a) efficiency, b) series resistance and c) characteristic time (lifetime).

Fig. 4. Current-voltage characteristics for an illuminated solar cell showing an initial increase in efficiency. *Inset:* Current-voltage characteristics near zero bias for the same solar cell.

Fig. 5. Efficiency (left axis) and dark zero-bias conductance (right axis) for a solar cell that exhibits a large initial efficiency increase before its exponential decrease.

a)
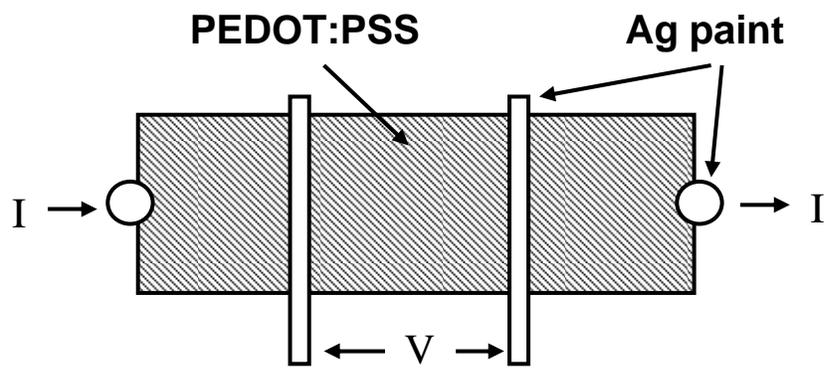

b)
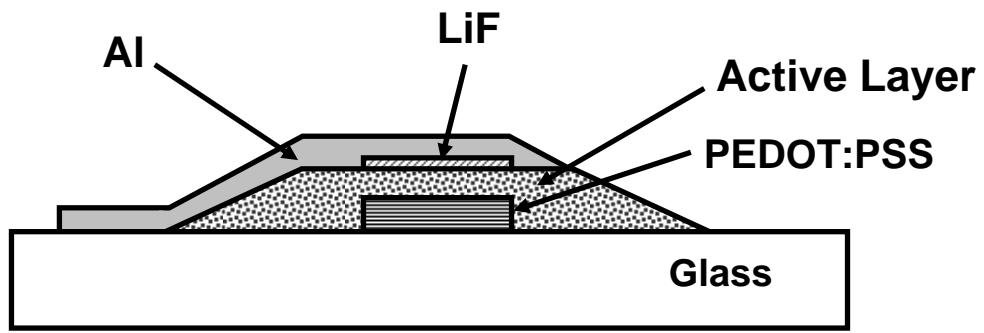

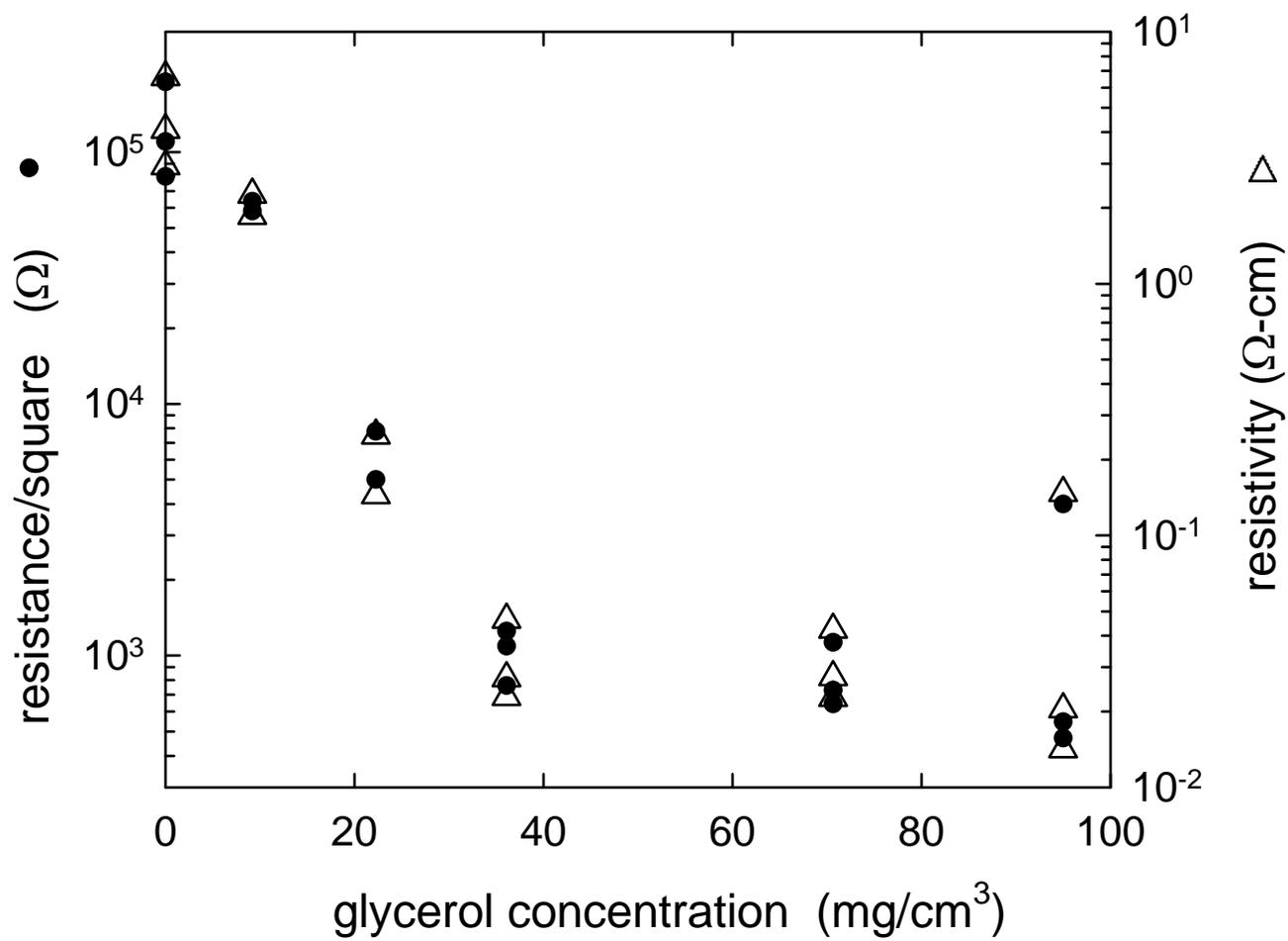

Fig. 2

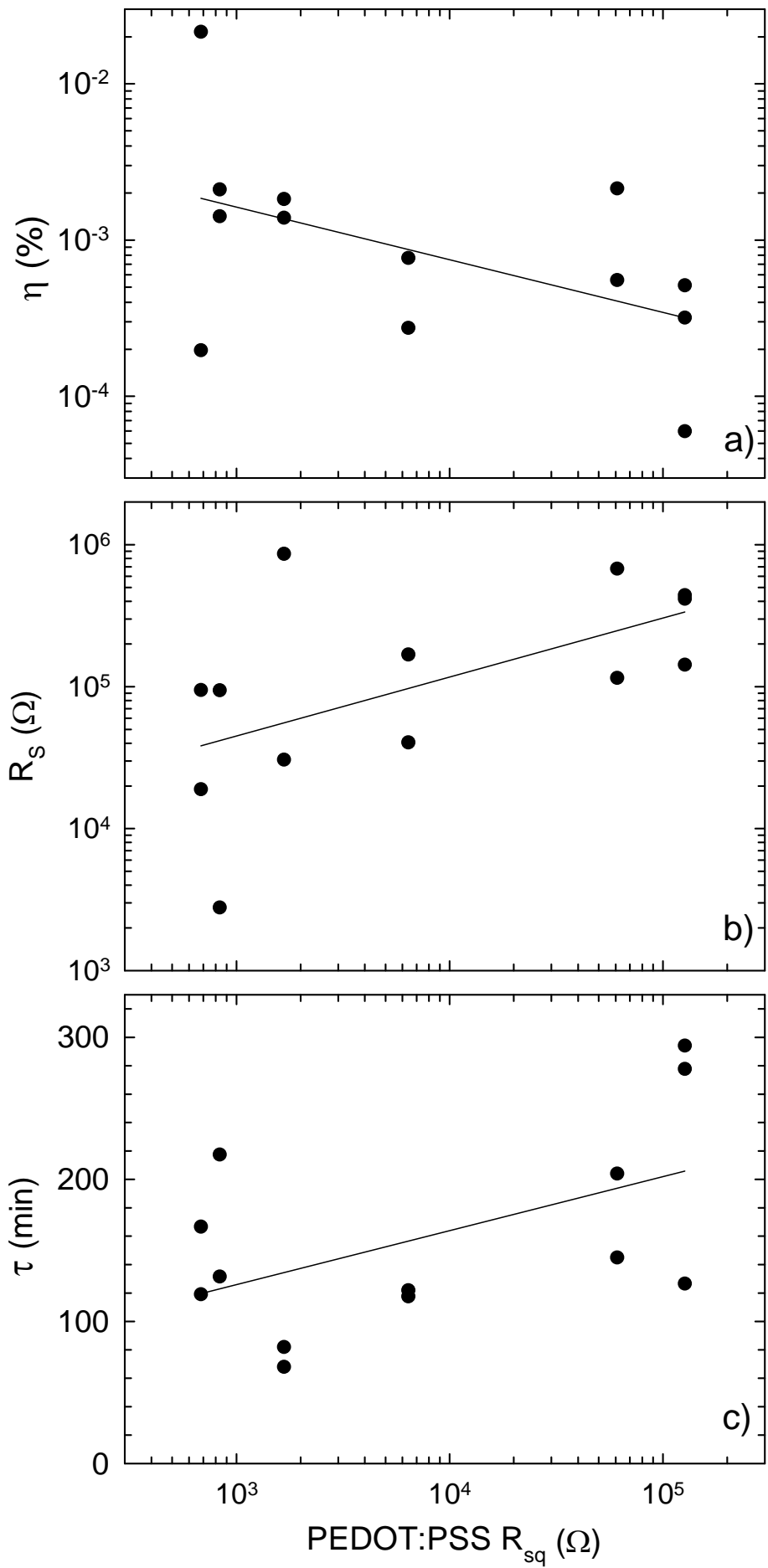

Fig. 3

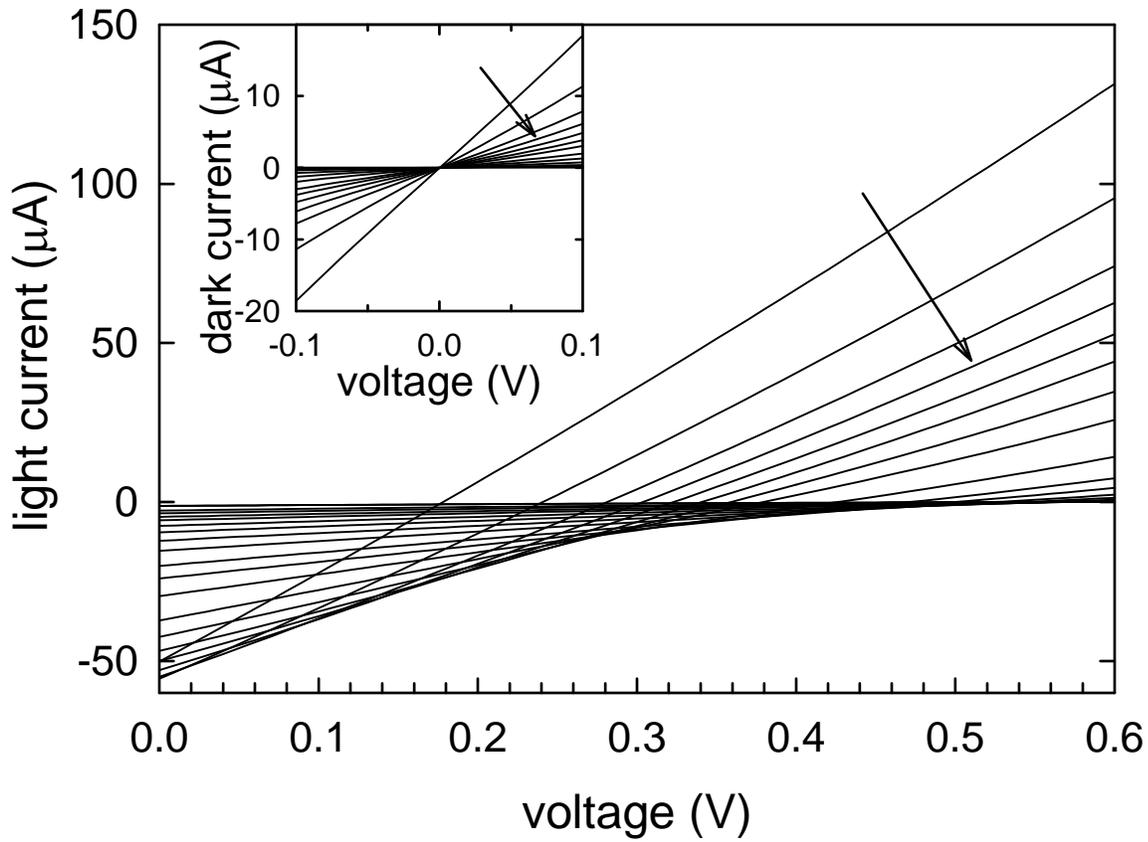

Fig. 4

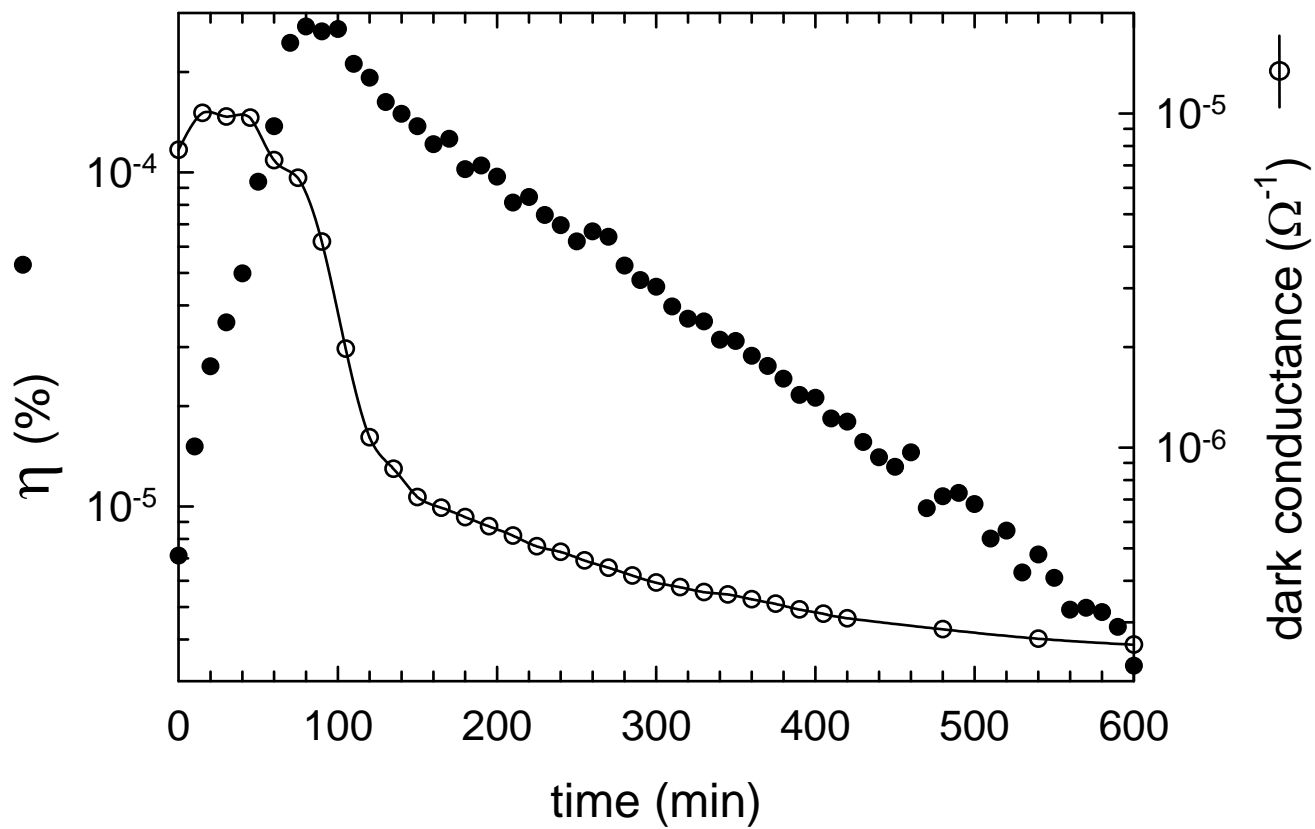

Fig. 5